# A new spin-functional MOSFET based on magnetic tunnel junction technology: pseudo-spin-MOSFET

Yusuke Shuto[1,5], Ryosho Nakane[2,5], Wenhong Wang[3,5], Hiroaki Sukegawa[3,5], Shuu'ichirou Yamamoto[4,5], Masaaki Tanaka[2,5], Koichiro Inomata[3,5], and Satoshi Sugahara[1,5]

[1]Imaging Science and Engineering Laboratory, Tokyo Institute of Technology, Yokohama 226-8502, Japan.
[2]Dept. of Electrical Engineering and Information Systems, The University of Tokyo, Tokyo 113-8656, Japan.
[3]Magnetic Materials Center, National Institute for Materials Science, Tsukuba 305-0047, Japan.
[4]Department of Information Processing, Tokyo Institute of Technology, Yokohama 226-8502, Japan.
[5]CREST, Japan Science and Technology Agency, Kawaguchi, 332-0012, Japan.

We fabricated and characterized a new spin-functional MOSFET referred to as a pseudo-spin-MOSFET (PS-MOSFET). The PS-MOSFET is a circuit using an ordinary MOSFET and magnetic tunnel junction (MTJ) for reproducing functions of spin-transistors. Device integration techniques for a bottom gate MOSFET using a silicon-on-insulator (SOI) substrate and for an MTJ with a full-Heusler alloy electrode and MgO tunnel barrier were developed. The fabricated PS-MOSFET exhibited high and low transconductance controlled by the magnetization configurations of the MTJ at room temperature. This is the first observation of spin-transistor operations for spin-functional MOSFETs.

*E-mail address: shuto@isl.titech.ac.jp

In recent years spin-transistors [1-6] have received considerable attention as a highly-functional building block of future integrated circuits [5-7].  In order to realize spin-transistors, it is essential that efficient spin injection/detection for their semiconductor channel is established [8-12].  However, this is not so easy in practice owing to several problems related to ferromagnet/semiconductor interface, such as interfacial layer formation, Fermi level pinning, and conductivity mismatch problems [8,12,13].  Although feasible technologies for these interface problems have been explored so far, crucial methods for them are still at the stage of searching.  We previously proposed a new circuit approach using an ordinary MOSFET and magnetic tunnel junction (MTJ) for reproducing functions of spin-transistors [14], referred to as a pseudo-spin-MOSFET (PS-MOSFET).  Recently the MTJ technology has dramatically progressed due to the development of MgO tunnel barriers.  Furthermore, half-metallic ferromagnet electrodes using full-Heusler alloys would also have a great impact on the MTJ technology.  In these situations, recently developed MTJs can exhibit higher tunneling magnetoresistance (TMR) ratios than 100% at room temperature [15-19], which is sufficient for spin-transistor operations of the PS-MOSFET and its applications to functional circuits such as nonvolatile logic circuitries [14, 20].  In this letter, we present fabrication of a PS-MOSFET and investigation of its spin-transistor operations that are controlled by the magnetization configurations of the MTJ connected to the PS-MOSFET.

Figure 1(a) shows the circuit configuration of the proposed PS-MOSFET.  The MTJ connected to the source terminal of the ordinary MOSFET feeds back its voltage drop to the gate, and the degree of the negative feedback depends on the resistance states of the MTJ. Therefore, effective input bias $V_{GS0}$ and also substrate (body-source) bias $V_{BS0}$ (shown in the figure) can be varied by the magnetization configurations of the MTJ even under a constant gate bias ($V_G$) condition. Therefore, the PS-MOSFET can possess high and low current drivabilities that are controlled by the magnetization configurations of the MTJ. In addition, magnetic-field-free current-induced magnetization switching (CIMS) [21] for the MTJ is also applicable. This can be performed when the drain current of the PS-MOSFET exceeds critical current for CIMS [14]. It should be noted that the transistor operation mode and CIMS mode of the PS-MOSFET can be separated by the amount of $V_G$. Therefore, the PS-MOSFET can reproduce the spin-transistor behaviors and would be the most promising spin-transistor based on the recently developed MTJ technology.

We developed integration technology of a spin-valve-type MTJ and bottom gate MOSFET for fabrication of a PS-MOSFET. Figure 1(b) shows the schematic side view of the fabricated PS-MOSFET. The simple bottom-gate structure using a silicon-on-insulator (SOI) substrate was employed for the MOSFET, in which the buried oxide (BOX) and Si substrate were used as a gate dielectrics layer and gate electrode, respectively. An SOI substrate with a p-type 100-nm-thick (001) SOI layer with resistivity of ~ 10 Ω·cm was used for the fabrication. The fabrication procedure was as follows: The SOI thickness was reduced from 100 nm to 20 nm by thermal oxidation of the SOI layer. The channel and source/drain regions were defined by the $SiO_2$ layer patterned as a hard mask, and then n-type impurity (phosphorus) was thermally doped into the source/drain regions. The physical channel length and width of the bottom gate MOSFET were 2 μm and 110 μm, respectively.

The spin-valve-type MTJ with a half-metallic full-Heusler alloy ($Co_2FeAl$; CFA) electrode and MgO tunnel barrier was fabricated on the thermally-grown atomically-flat $SiO_2$ layer adjacent to the source region. Firstly, a Ru(7nm) / IrMn(12nm) / CoFe(3nm) / MgO(1.5nm) / CFA(20nm) multilayer was deposited by RF sputtering at room temperature on the $SiO_2$ layer using a 10-nm-thick MgO buffer layer. Recently it was confirmed that MTJs with full-Heusler alloy electrodes formed on an amorphous SiO$_2$ layer exhibited relatively high TMR ratios, using a highly-oriented MgO buffer layer [22]. Subsequently, a post-annealing treatment was performed at 300°C for quality improvement of the CFA film. After this treatment, the CFA film exhibited a highly (001)-oriented *B*2-ordered structure. During the post-annealing, a magnetic field was applied to the sample for sufficient exchange biasing between the CoFe and IrMn layers. Then, the multilayer film was formed into a rectangular shape of 15 × 50 μm$^2$. The source region of the MOSFET was connected to the bottom electrode of the MTJ using an Al interconnect, and also Al pads for the drain contact of the MOFET and the top electrode of the MTJ were formed. Finally, AuGa was deposited on the back side of the Si substrate as a contact pad for the bottom gate. Figure 1(c) shows the photograph of the fabricated PS-MOSFET.

In our preceding study, a prototype PS-MOSFET was fabricated [23]. Owing to its severe gate and source/drain leakage currents, only primitive results for magnetization-configuration-dependent output characteristics were observed. Therefore, the fabrication process conditions were reexamined and refined. The bottom-gate MOSFET fabricated in this study exhibited clear field-effect transistor characteristics, although it showed depletion mode behavior and slight leakage currents through the source junction and gate dielectrics layer (BOX). The former would be caused by the heavy diffusion process of P atoms to form the low-resistive source/drain regions, and the latter would be caused by ion milling process to form the MTJ pillar. Although more process condition improvement would be required, magnetocurrent characteristics of the PS-MOSFET were successfully evaluated, as discussed in this letter.

Figures 2(a) and (b) show the electrical characteristics of the MTJ in the fabricated PS-MOSFET at room temperature (RT). The MTJ exhibited the clear exchange-biased TMR behavior and a relatively high TMR ratio of 71.3% at RT, as shown in Fig. 2(a). The resistance ($R_P$) in parallel magnetization measured with a bias voltage of 10 mV was 316 Ω that was consistent with a resistance-area product ($RA$) value of 237 kΩ·μm². The characteristic voltage so-called $V_{half}$ (that is a bias voltage when the TMR ratio was reduced to half its original value) was 0.64V.

Figure 3 shows the output characteristics of the fabricated PS-MOSFET. The drain currents are plotted as a function of drain bias ($V_D$) swept from 0 to 2V, where the gate bias ($V_G$) varies from -2 to 3 V in steps of 1V. Solid curves ($I_D^P$) and broken curves ($I_D^{AP}$) in the figure show the drain currents in the parallel and antiparallel magnetization configurations, respectively. The depletion-type field-effect transistor behavior was clearly observed, as described previously. $I_D^P$ was higher than $I_D^{AP}$ over the entire linear and saturation regions, indicating that the PS-MOSFET successfully operated as a spin-transistor. Figure 4(a) shows the drain current of the PS-MOSFET as a function of magnetic field at RT, where $V_D$ = 100mV and $V_G$ = 2.0V. The behavior of the drain current well reflected the resistance change of the MTJ shown in Fig. 2(a). The magnetocurrent ratio ($\gamma_{MC}$) defined as $\gamma_{MC} = (I_D^P - I_D^{AP})/I_D^{AP}$ was 38.4% at $V_D$ = 0.1V and $V_G$ = 2V. Solid curves in Fig. 4(b) show $\gamma_{MC}$ as a function of $V_D$ at RT, where $V_G$ varies from 0 to 5 V in steps of 1V. $\gamma_{MC}$ increased with decreasing $V_D$ and increased with increasing $V_G$. Dashed curves in Fig. 4(b) show calculated magnetocurrent ratio $\gamma_{MC}^{cal}$ using SPICE program with our developed MTJ macromodel [24]. The simulation was able well to reproduce the output characteristics of the fabricated PS-MOSFET except the leakage current components. The $V_D$-dependence of $\gamma_{MC}$ was consistent with $\gamma_{MC}^{cal}$ for a range of $V_G < \sim 2V$. However, $\gamma_{MC}$ exceeded $\gamma_{MC}^{cal}$ when $V_G$ was higher than 3V. This can be attributed to the gate leakage current of the bottom-gate

MOSFET. $I_D^P$ and $I_D^{AP}$ at the higher $V_G$ conditions (> 3V) decreased owing to the gate leakage current, and the effect of the leakage current comes to stand out more on a condition of lower $V_D$ (<~ 0.7V). In particular highly reduced $I_D^{AP}$ enhanced significantly $\gamma_{MC}$. Therefore, the steep increase in $\gamma_{MC}$ more than $\gamma_{MC}^{cal}$ shown in Fig. 4(b) is apparent, and $\gamma_{MC}$ would take a maximum value of ~ 45 %, when the gate leakage current is diminished. Note that the enhancement of $\gamma_{MC}$ (not apparent component, see $\gamma_{MC}^{cal}$) at lower $V_D$ and higher $V_G$ conditions are useful for the operation of nonvolatile SRAM and nonvolatile flip-flop (FF) circuits [14, 20] that are important applications for PS-MOSFETs.

In the application of PS-MOSFETs to NV-SRAM and NV-FF circuits, a moderate TMR ratio of ~ 100 % is sufficient (i.e., an extremely high TMR ratio is not required) [14]. This is due to the nature of a bistable circuitry used in the NV-SRAM and NV-FF circuits [14, 20]. In this letter, we used the MTJ with the TMR ratio of 71.3 % that was slightly lower than the required value for the NV-SRAM/NV-FF application, and this MTJ was formed on a thermally-grown amorphous $SiO_2$ layer. In actual applications of PS-MOSFETs to integrated circuits, it is necessary to fabricate MTJs on an interlayer dielectrics ($SiO_2$ or related silica) layer in multilevel interconnections. Recently it was reported that MTJs using full-Heusler $Co_2FeAl_{0.5}Si_{0.5}$ alloy electrodes formed on a self-crystallized MgO buffer layer exhibited higher TMR ratios than 100 % even on a $SiO_2$/Si substrate [22]. Moreover, interface crystallization of amorphous CoFeB electrodes in CoFeB/MgO/CoFeB trilayer structures is also effective to achieve a high TMR ratio for MTJs formed on an amorphous $SiO_2$ layer [25, 26]. These techniques would be applicable to fabrication of such high-TMR-MTJs on an interlayer dielectrics layer of integrated circuits. Preparation of an atomically flat and smooth surface of an interlayer dielectrics layer (or metal layer formed on it) would be an important issue for these approaches. Therefore, we can expect that PS-MOSFETs would be the most promising spin-transistor that can be achieved by diverting the MTJ technology.

In summary, we fabricated and characterized a PS-MOSFET that is a new circuit for reproducing functions of spin-transistors using an ordinary MOSFET and MTJ. Device integration technique for a bottom gate MOSFET using an SOI substrate and for an MTJ with a full-Heusler alloy electrode and MgO tunnel barrier were developed. The fabricated PS-MOSFET exhibited high and low transconductance controlled by the magnetization configurations of the MTJ at room temperature. Our proposed PS-MOSFET would be the most promising spin transistor based on the MTJ technology.

**Acknowledgements**

This work was supported by Core Research for Evolutional Science and Technology (CREST) of Japan Science and Technology Agency (JST). The authors would like to thank Prof. H. Munekata, Tokyo Institute of Technology.

**Figure captions**

FIG. 1: (a) Circuit configuration of the proposed pseudo-spin-MOSFET (PS-MOSFET). (b) Schematic side view of the fabricated PS-MOSFET. (c) Photograph of the fabricated PS-MOSFET.

FIG. 2: (a) Junction resistance of the fabricated MTJ as a function of magnetic field at room temperature (300K). (b) Junction resistance as a function of applied voltage for the MTJ in the parallel and antiparallel magnetization configurations.

FIG. 3: Output characteristics of the fabricated PS-MOSFET at room temperature. The drain currents are plotted as a function of drain bias $V_D$, where gate bias $V_G$ varies from -2 to 3V in steps of 1V. Solid curves ($I_D^P$) and broken curves ($I_D^{AP}$) show the drain currents in the parallel and antiparallel magnetization configurations, respectively.

FIG. 4: (a) Drain current as a function of magnetic field at room temperature, measured with $V_D = 0.1V$ and $V_G = 2V$. (b) Magnetocurrent ratio $\gamma_{MC}$ as a function of $V_D$ at room temperature, where $V_G$ varies from 0 to 5V in steps of 1V. Dash curve shows calcurated magnetocurrent ratio $\gamma_{MC}^{cal}$.

**Figures**

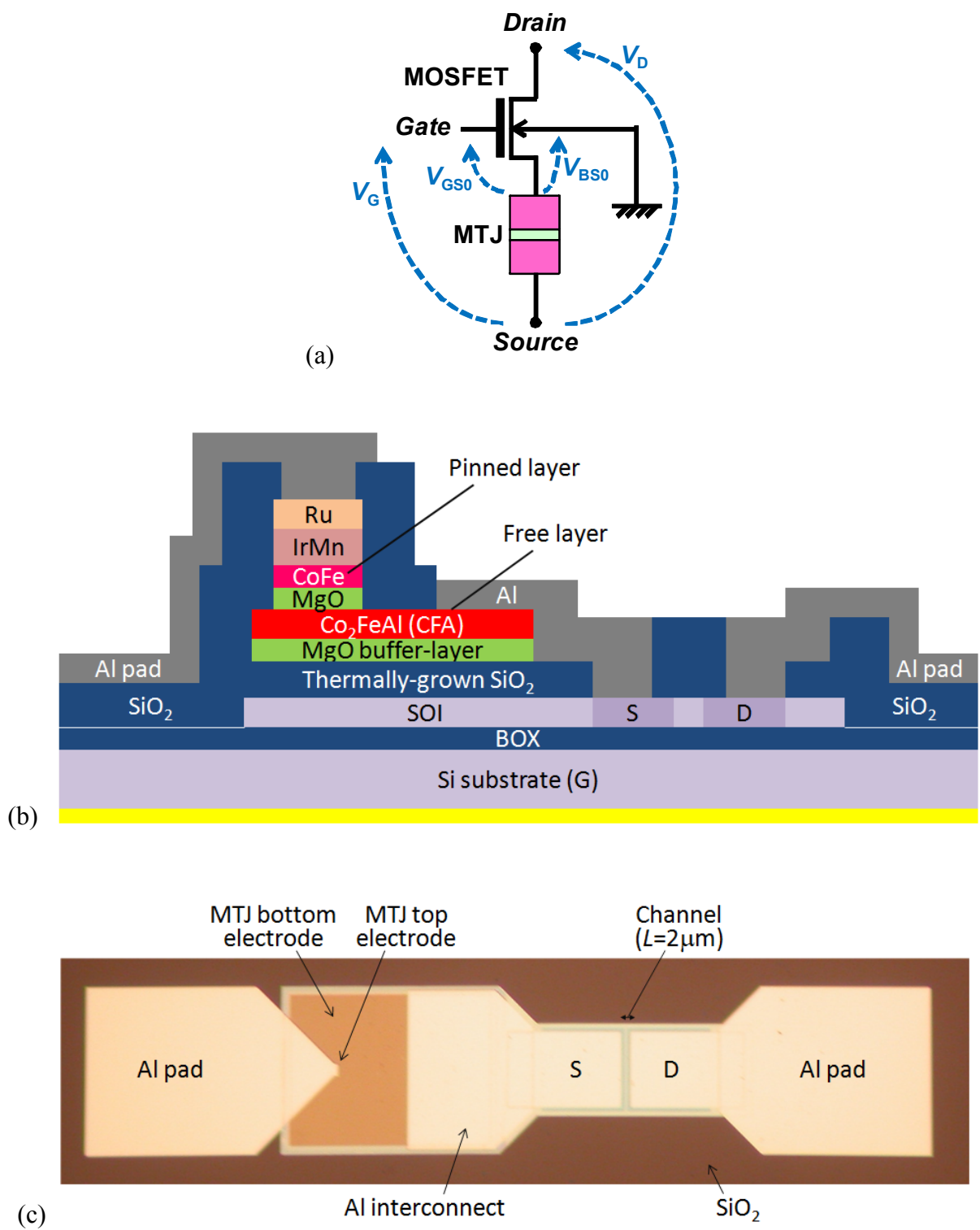

Figure 1:   Shuto *et al*.

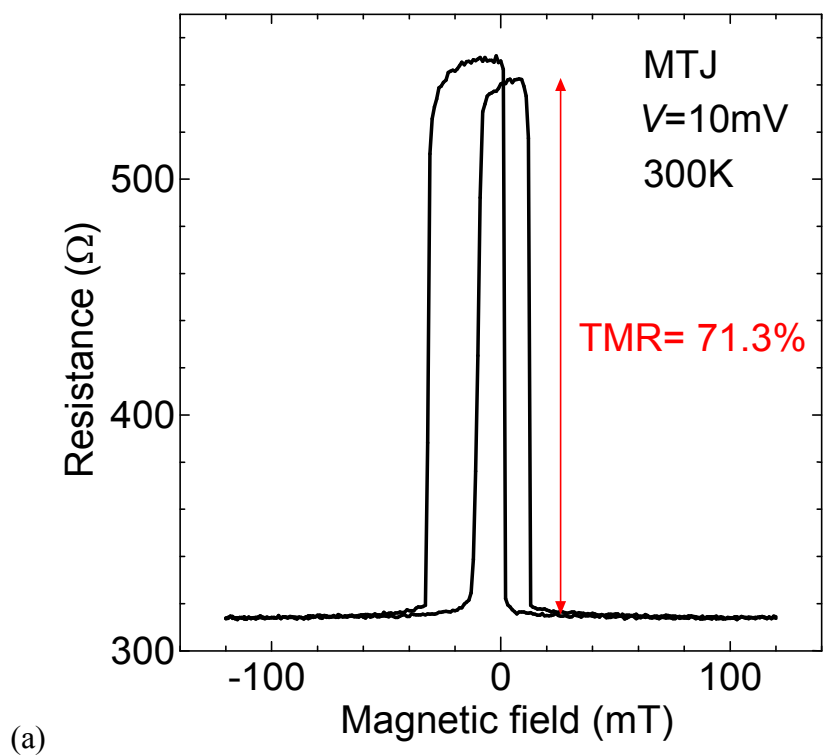

(a)

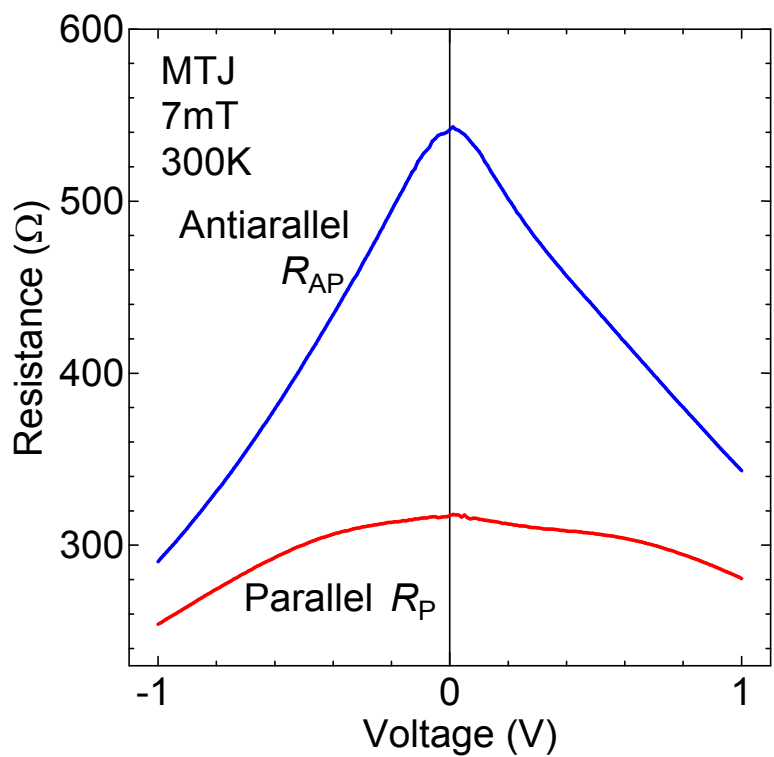

(b)

Figure 2: Shuto *et al*.

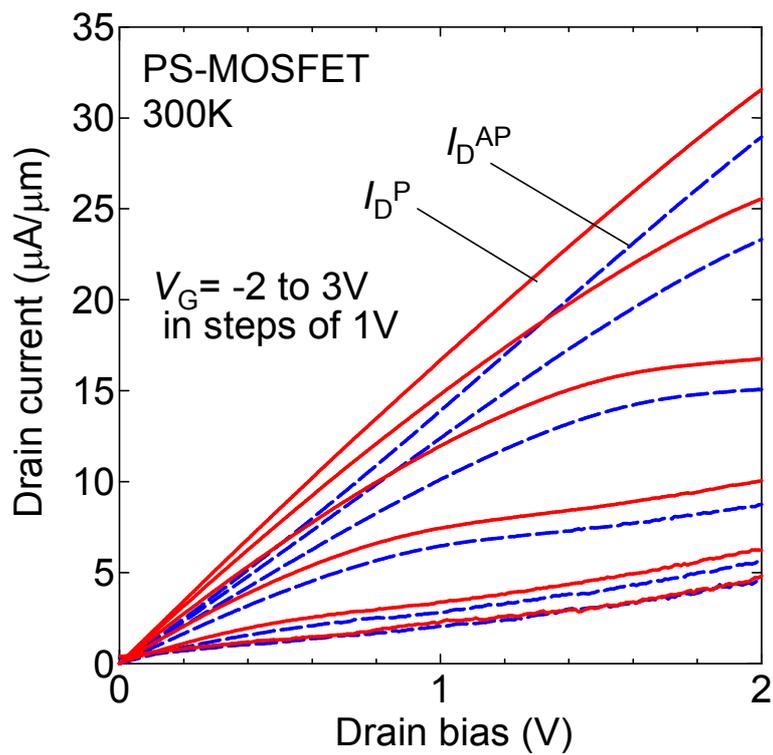

Figure 3:   Shuto *et al*.

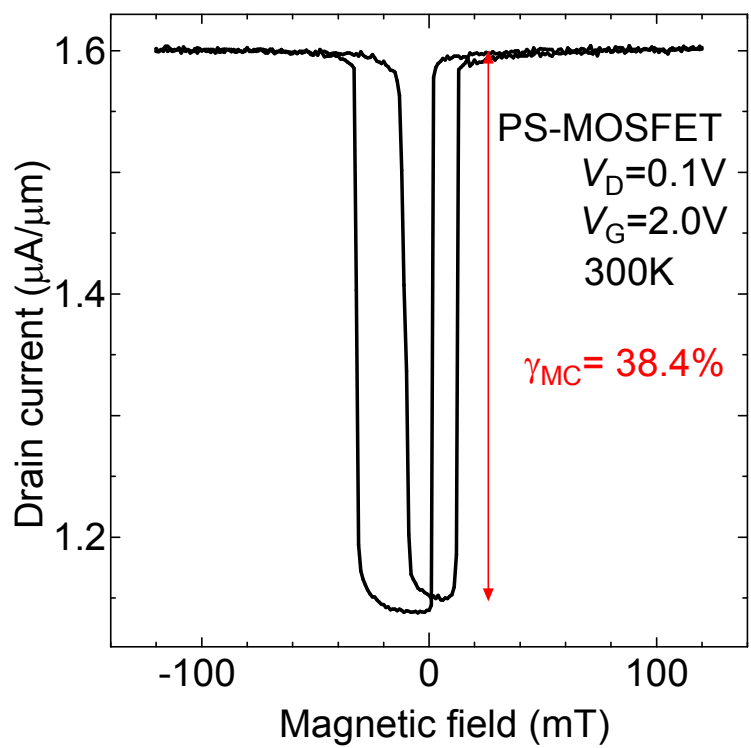

(a)

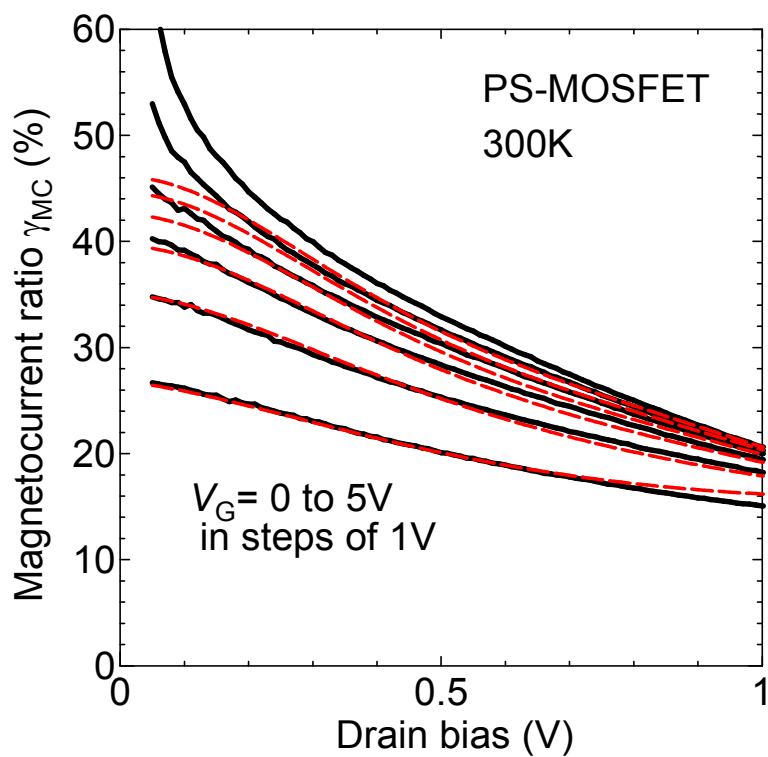

(b)

Figure 4: Shuto *et al*.